\documentclass[twocolumn,notitlepage,longbibliography]{revtex4-1}
\usepackage{amsmath}
\usepackage{amssymb}
\usepackage{bm}
\usepackage{epsfig}
\usepackage{graphicx}
\usepackage{color}
\usepackage[T2A]{fontenc}
 \usepackage[cp1251]{inputenc}
 \usepackage[russian,english]{babel}
\usepackage[unicode=true,colorlinks=true,citecolor=blue]{hyperref}
\usepackage{enumitem}

\usepackage{rotating,subfigure}
\usepackage[3D]{movie15}

\renewcommand{\Re}{\mathop{\rm Re}}

\renewcommand{\i}{\mathrm{i}}

\newcount\timehh  \newcount\timemm
\timehh=\time \divide\timehh by 60
\timemm=\time
\begin{document}
\title{Stochastic Faraday rotation induced by the electric current fluctuations\\ in nanosystems}

\author{D. S. Smirnov}
\affiliation{Ioffe Institute, 194021,
St.-Petersburg, Russia}
\author{M. M. Glazov}
\affiliation{Ioffe Institute, 194021,
St.-Petersburg, Russia}

\begin{abstract}
We demonstrate theoretically that in gyrotropic semiconductors and semiconductor nanosystems the Brownian motion of electrons results in temporal fluctuations of the polarization plane of light passing through or reflected from the structure, i.e., in stochastic Faraday or Kerr rotation effects. The theory of the effects is developed for a number of prominent gyrotropic systems such as bulk tellurium, ensembles of chiral carbon nanotubes, and GaAs-based quantum wells of different crystallographic orientations. We show that the power spectrum of these fluctuations in thermal equilibrium is proportional to the \emph{ac} conductivity of the system. We evaluate contributions resulting from the fluctuations of the electric current, as well as of spin, valley polarization, and the spin current to the noise of the Faraday/Kerr rotation. Hence, all-optical measurements of the Faraday and Kerr rotation noise provide an access to the transport properties of the semiconductor systems.
\end{abstract}

\maketitle

\section{Introduction}\label{sec:intro}

In gyrotropic systems certain components of  polar vectors, such as light wavevector, $\bm Q$, or electric current density, $\bm j$, and of axial vectors (pseudovectors), e.g., magnetic field, $\bm B$, transform in a same way. Such a symmetry wise indistinguishability between the components of polar and axial vectors manifests itself in a variety of effects, including circular photogalvanic effect~\cite{sturmanBOOK,ivchenko_pikus_problems_eng,ivchenko1978new}, spin orientation by electric current~\cite{aronov89:eng,edelstein90,PhysRevLett.93.176601,silov04,Ganichev2006127} and spin galvanic effect~\cite{ivchenko1989photocurrent,ganichev02}, linear in the electron wavevector splittings of energy bands~\cite{rashba64,golub_ganichev_BIASIA}, etc. One of the most prominent examples of such phenomena is the natural optical activity, the rotation of the plane of polarization of linearly polarized light as it travels through a gyrotropic medium~\cite{ll8_eng}. Back in 1970s it was predicted that a current flowing through a gyrotropic medium results in the light polarization plane rotation~\cite{Baranova1977243,ivchenko1978new}, the effect was observed in bulk Te shortly afterwards~\cite{vorob1979optical}.

In this paper we point out that omnipresent stochastic fluctuations of electric current, $\delta \bm j$, resulting, e.g., from the Brownian motion of electrons in the thermal equilibrium give rise to the stochastic rotation of the light polarization plane as it passes through a gyrotropic system. Previously, the fluctuations of the Faraday or Kerr rotation of the light polarization plane transmitted through or reflected from the medium have been related with the spin fluctuations in the system~\cite{aleksandrov81}. This has led to the development of the spin noise spectroscopy technique widely used nowadays to study the spin dynamics of atoms, electrons and nuclei~\cite{Crooker_Noise,Oestreich:rev,Zapasskii:13,2015arXiv150605370B}. Here we demonstrate that, in addition to the spin noise, in gyrotropic systems, the spectrum of the Faraday and Kerr rotation fluctuations contains information about the electric current fluctuations. 

In Sec.~\ref{sec:phen} we provide the phenomenological description of the effect based on the symmetry arguments and fluctuation-dissipation theorem. Further, in Sec.~\ref{sec:micro} we develop the microscopic theory of the effect for the most prominent gyrotropic semiconductor systems: bulk Te, chiral carbon nanotubes, and quantum wells grown from the zincblende lattice semiconductors such as GaAs. The results are discussed in Sec.~\ref{sec:discussion} and summarized in Sec.~\ref{sec:conclusion}.

\section{Phenomenological description}\label{sec:phen}

\begin{figure}
\includegraphics[width=\linewidth]{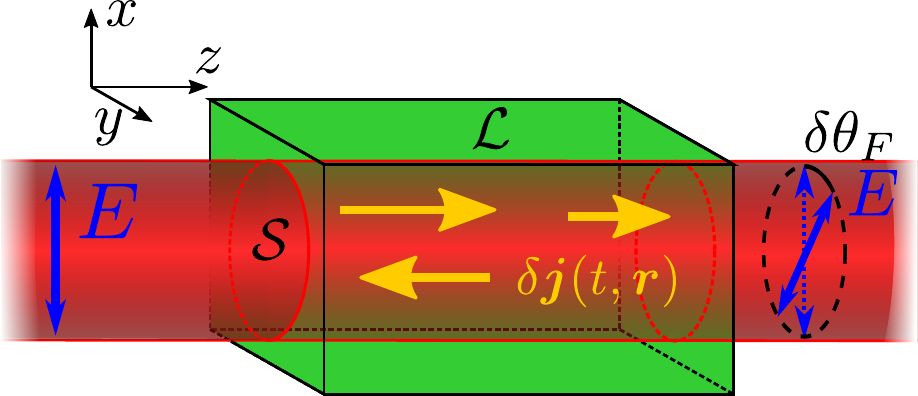}
\caption{Schematic illustration of the studied system. The fluctuations of electric current in the sample lead to the stochastic Faraday rotation of the probe beam.} \label{fig:scheme}
\end{figure}

Let us consider a gyrotropic medium described by the dielectric permittivity tensor with the Cartesian components $\varepsilon_{\alpha\beta}$, hereafter greek subscripts $\alpha,\beta,\gamma,\ldots$ run through $x,y,z$. Such a description is applicable to bulk semiconductors, multiple quantum well structures and ensembles of nanotubes~\cite{ivchenko05a}. We assume that there are free charge carriers in the system. The optical activity induced by the macroscopic current with the density $\bm j$ is described by the phenomenological relation $\varepsilon_{\alpha\beta} = \mathrm i p_{\alpha\beta\gamma} j_\gamma$, where $p_{\alpha\beta\gamma} = - p_{\beta\alpha\gamma}$ is the third rank tensor asymmetric with respect to the permutation of the first two subscripts~\cite{ivchenko1978new,vorob1979optical}.
Hereafter the summation over the repeated Greek subscripts, i.e., over the $\gamma$, is assumed. 
Due to the Brownian motion of the charge carriers the electric current fluctuates in time and space giving rise to stochastic time and coordinate dependent fluctuations of the current density $\delta \bm j=\delta \bm j( t, \bm r)$. These current fluctuations, in turn, give rise to the fluctuations of the off diagonal components of the dielectric tensor
\begin{equation}
\label{fluct}
\delta \varepsilon_{\alpha\beta}(t,\bm r) = - \delta \varepsilon_{\beta\alpha}(t, \bm r) = \mathrm i p_{\alpha\beta\gamma} \delta j_\gamma(t, \bm r).
\end{equation}
Therefore, the current fluctuations lead to fluctuating circular birefringence and dichroism. The fluctuations can thus can be probed by measuring the noise of the Faraday rotation angle $\delta \theta_F$ of the polarization plane for a linearly polarized light propagating through a sample with the wavevector $\bm Q$ for specificity or of the Kerr rotation angle $\delta\theta_K$ of the polarization plane of light reflected from the sample. To be specific, we consider below the case of the Faraday effect and $\bm Q\parallel z$ where the rotation angle fluctuation can be recast as 
\begin{equation}
\label{fluct:F}
\delta \theta_F(t) = \frac{\omega}{2cn}  p_{yx\gamma}\int \frac{d\bm r}{\mathcal S} \delta j_\gamma(t, \bm r).
\end{equation}
Here $\omega$ is the frequency of light, $c$ is the speed of light, $n$ is the background refraction index, and $\mathcal S$ is the area of illuminated spot in the sample, Fig.~\ref{fig:scheme}. The anisotropy of the background dielectric constants is disregarded. Equation~\eqref{fluct:F} is similar to the well known expression relating regular or stochastic Faraday rotation angle with the magnetization in the illuminated volume~\cite{ll8_eng,aronovivchenko:eng,PhysRevB.85.195313}. It can be rigorously derived assuming that the incident field is linearly polarized, e.g., along the $x$-axis and expressing the induced $y$-component of the electric field via the electrodynamical Greens function, $G_{yy}(\bm r)$, as $\delta E_y = (\omega/c)^2\int G_{yy}(\bm r- \bm r') \delta \varepsilon_{yx}(\bm r')E_x(\bm r') d\bm r'$. This expression is valid provided that the timescale of the dielectric tensor variations (which is the time scale of the current fluctuations) is much longer as compared with the period of the electromagnetic field oscillations and the time of light passage through the sample. The measured Faraday rotation signal can be recast in the form
 $\delta\theta_F= (|E_x|^2 S)^{-1}\int d\bm \rho \Re\{\delta E_yE_x^*\}$~\cite{glazov:review}, where
the integration is carried out over the detector surface area, $S$.
 These expressions together with the asymptotic form of the Greens function, $G_{\alpha\alpha}(\bm r) = \exp{(\mathrm i Q r)}/(4\pi r)$, yield Eq.~\eqref{fluct:F}. Note, that the Faraday rotation caused by the natural optical activity of the medium, $\theta_F \propto Q$, is static and, therefore, does not contribute in the noise of the rotation signal~\eqref{fluct:F} detected in experiments.

In the Faraday rotation noise experiments the Fourier transform of the Faraday rotation autocorrelation function,
\begin{equation}
\label{correlation}
(\delta \theta_F^2)_\Omega = \int_{-\infty}^\infty \langle \delta\theta_F(t+\tau)\delta\theta_F(t)\rangle e^{\mathrm i \Omega \tau} d\tau,
\end{equation}
is measured Here and in what follows we consider only equilibrium fluctuations, hence, the averaging over $t$ in the definition of the autocorrelation function, Eq.~\eqref{correlation}, is assumed. Making use of Eq.~\eqref{fluct:F} we finally obtain
\begin{equation}
\label{corr:F}
(\delta \theta_F^2)_\Omega = \left(\frac{\omega{\mathcal{L}}}{2cn}\right)^2  p_{yx\gamma}p_{yx\gamma'} \left(\delta j_\gamma \delta j_{\gamma'} \right)_{\Omega}^{\bm q=0},
\end{equation}
where $\mathcal{L}$ is the length of the sample, $\left(\delta j_\gamma \delta j_{\gamma'} \right)_{\Omega}^{\bm q=0}$ is the Fourier transform of the correlation function of homogeneous in the space ($\bm q=0$) fluctuations of electric current density components in the illuminated volume of the sample, $\mathcal{V}=\mathcal{LS}$. For completeness we present here the definition of the current density components correlation functions~\cite{springerlink:10.1007/BF02724353}
\begin{multline}
\left(\delta j_\gamma \delta j_{\gamma'} \right)_{\Omega}^{\bm q} =\\ \frac{1}{\mathcal V} \int \langle \delta j_\gamma(t+\tau, \bm r + \bm r_1 )\delta j_{\gamma'} (t, \bm r)\rangle  e^{\mathrm i \Omega \tau - \mathrm i \bm q \cdot \bm r_1} d\tau d\bm r_1,
\nonumber
\end{multline}
with the angular brackets denoting averaging over $t$ and $\bm r$. In what follows the superscript $\bm q=0$ is omitted.

In the thermal equilibrium the correlator $\left(\delta j_\gamma \delta j_{\gamma'} \right)_{\Omega}$ in Eq.~\eqref{corr:F} is proportional to the \emph{ac} conductivity of the system, $\sigma_{\gamma\gamma'}(\Omega)$,
\begin{equation}
\label{conductivity}
\left(\delta j_\gamma \delta j_{\gamma'} \right)_{\Omega} = \frac{2k_B T}{\mathcal V} \Re\, \sigma_{\gamma\gamma'}(\Omega),
\end{equation}
where $T$ is the temperature and $k_B$ is the Boltzmann constant~\cite{0034-4885-29-1-306,springerlink:10.1007/BF02724353}.
In the nonequilibrium steady state conditions the correlators of current fluctuations do not, in general, reduce to any response function, and can be calculated following the formalism proposed in Ref.~\onlinecite{springerlink:10.1007/BF02724353}.

Equations~\eqref{corr:F} and~\eqref{conductivity} can be combined as 
  \begin{equation}
    (\delta \theta_F^2)_\Omega = \frac{2k_B T  \mathcal{L}}{\mathcal S} \left(\frac{\omega}{2cn}\right)^2 p_{yx\gamma}p_{yx\gamma'} \Re\, \sigma_{\gamma\gamma'}(\Omega),
\label{eq:Faraday_sigma}
  \end{equation}
which clearly demonstrates that the electric current fluctuations described by the conductivity of the system manifest themselves in the noise of the Faraday rotation. Particularly for the conductivity described by the Drude formula~\cite{Note2,springerlink:10.1007/BF02724353,platzman1973solid}.
\begin{equation}
 \sigma_{\gamma\gamma'}(\Omega)=\frac{Ne^2\tau}{m(1-\i\Omega\tau)}\delta_{\gamma\gamma'},
  \label{eq:Drude}
\end{equation}
where $N$ is the concentration of free charge carriers, $m$ is the charge carrier effective mass, $\tau$ is its momentum relaxation time, and $\delta_{\gamma\gamma'}$ is the Kronecker symbol, one can readily obtain the expression for the mean square fluctuation of Faraday rotation angle
\begin{equation}
  \left\langle\delta\theta_F^2\right\rangle=\int_{-\infty}^\infty (\delta \theta_F^2)_\Omega \frac{d\Omega}{2\pi}=
  \frac{Ne^2k_BT}{m} \frac{\mathcal L}{\mathcal S}\sum_\gamma\left(\frac{\omega p_{yx\gamma}}{2cn}\right)^2.
  \label{eq:RMS}
\end{equation}
The magnitude of the single time mean square fluctuation of the Faraday rotation angle is, in agreement with physical arguments, linearly proportional to the charge carrier density and the temperature and is inversely proportional to the effective mass. We note that the measured signal scales as $\mathcal{L}/\mathcal{S}$, which paves the way towards increasing of the sensitivity by increasing the sample length and making the focussing of the probe beam tighter just like in the conventional spin noise spectroscopy~\cite{Oestreich:rev,Glazov:15}.

In the following section we provide the microscopic theory for $p_{\alpha\beta\gamma}$ for a number of gyrotropic semiconductor systems and give the estimations of the root mean square Faraday rotation $\sqrt{\left\langle\delta\theta_F^2\right\rangle}$ in these systems.

\section{Microscopic models}\label{sec:micro}

Below we consider four prominent gyrotropic systems, namely, bulk tellurium, macroscopic ensemble of chiral carbon nanotubes, and quantum well structures based on GaAs (or similar) semiconductor grown along $[110]$ and $[001]$ axes.

\subsection{Bulk tellurium}\label{subsec:Te}

\begin{figure}
\begin{minipage}{\linewidth}
{\large (a)}~~~~~~~~~~~~~~~~~~~~~~~~~~~~~~~~~~~~~~~~~~~~~~~~~~~~~~~~~~~\vspace{-1cm}

~~~~~~\includemovie[poster,
			  toolbar,
	          label=pt,
	          text={\includegraphics[width=\linewidth]{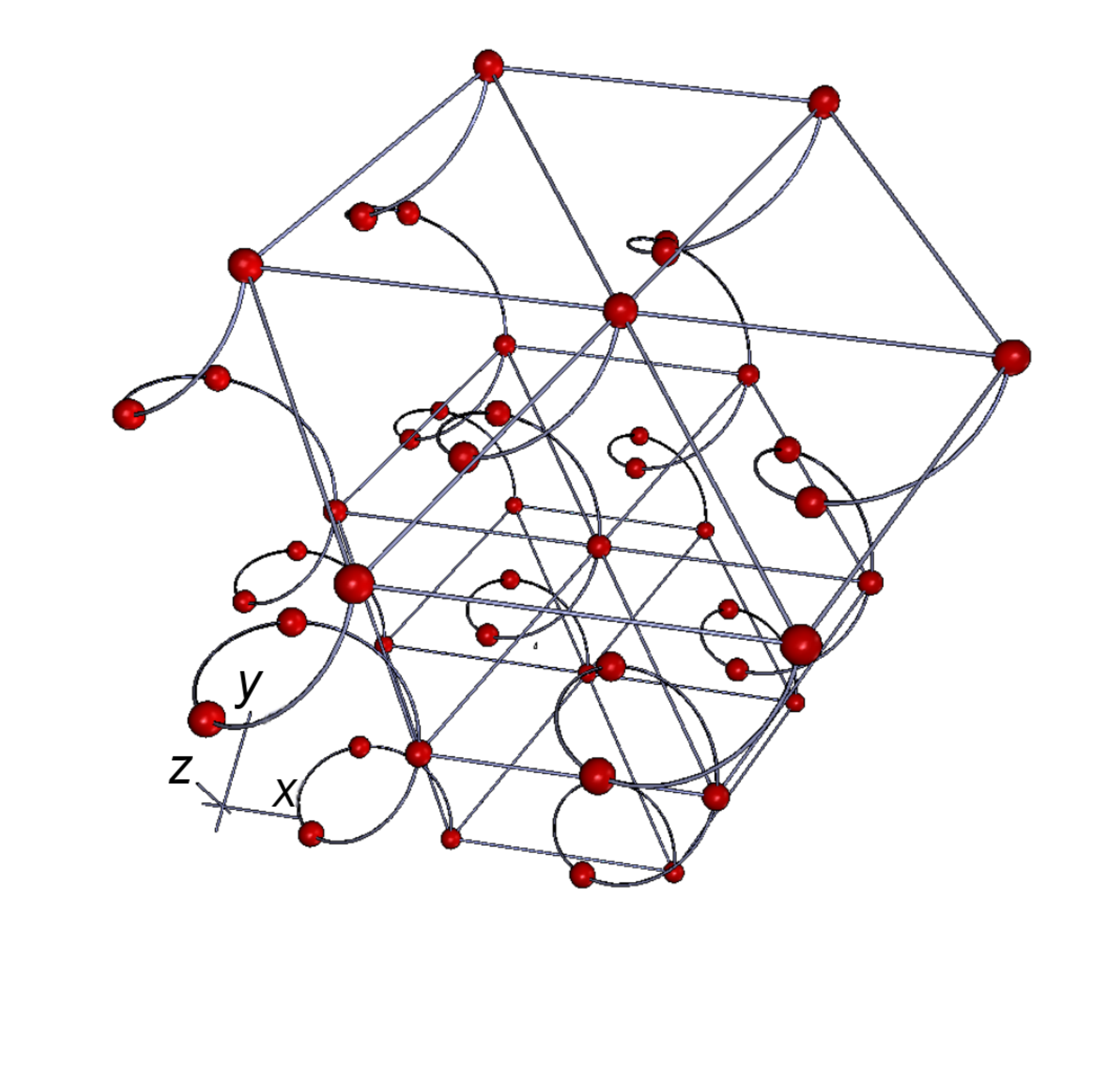}},
3Droo=1.7532683988025128,
3Daac=60.000001669652114,
3Dcoo=0.05479823797941208 0.000023365430024568923 0.5929998755455017,
3Dc2c=0.05754242092370987 -0.45956751704216003 0.8862767815589905,
3Droll=-1.2335996122223567,
3Dlights=CAD,
              3Drender=SolidOutline]
			  {\linewidth}{\linewidth}{helixXYZ.u3d}
\end{minipage}
\begin{minipage}{\linewidth}
{~~~~~~\large (b)\hfill}\vspace{-0.3cm}

\includegraphics[width=0.8\textwidth]{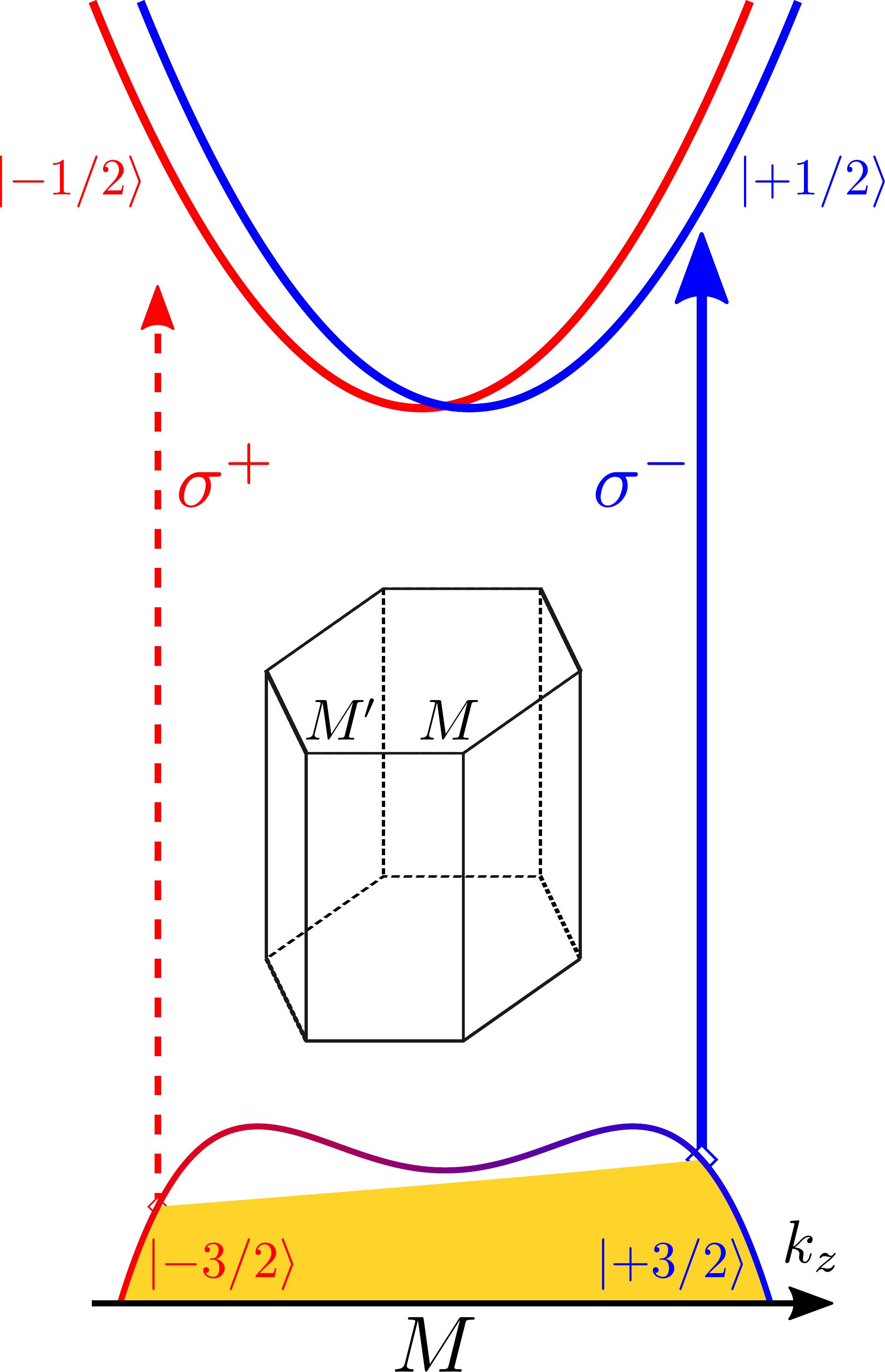}
\end{minipage}
  \caption{(a) (Interactive online)\footnote{In \href{https://acrobat.adobe.com/us/en/acrobat/pdf-reader.html}{Adobe Acrobat viewer} click the Options button and then select ``Trust this document'' option.}
The crystal structure of bulk right handed tellurium, balls show the atoms of Te. (b) Band structure of bulk $p$-doped Te in the vicinity of $M$ point, the filled area shows the occupied states in the presence of the current $j_z<0$.
Solid and dashed arrows show the transitions with smaller and larger detunings, respectively, whose imbalance results in the Faraday rotation.
The inset shows Te Brillouin zone with the equivalent extrema of the conduction and valence bands denoted as $M$ and $M'$.}
\label{fig:Te}
\end{figure}

The bulk Te has $D_3$ symmetry being a prototypical gyrotropic crystal~\cite{ivchenko1975natural,dubinskaya1978natural}, whose electrooptical properties attract continued interest~\cite{Agapito2013,PhysRevB.93.045207}. It has two modifications: right and left handed, in which the atoms are placed along the spirals with different chiralities~\cite{Tanaka2010}. The axes of the spirals form the hexagonal lattice, as sketched in Fig.~\ref{fig:Te}(a) for the example of right handed modification. The extrema of the conduction and valence bands are located near $M$ and $M'$ points of the Brillouin zone [see inset in Fig.~\ref{fig:Te}(b)], and the band structure in the vicinity of these extrema points is illustrated in Fig.~\ref{fig:Te}(b). The conduction band is twofold spin degenerate at $M$ and $M'$ points. The dispersion for electrons with spin projection on the threefold rotation axis $z$, $s=\pm1/2$, is given by~\cite{ivchenko1978new}
\begin{equation}
  {E_c^{s}}= E_c + \beta_ck_zs = E_g+ \frac{\hbar^2k_z^2}{2m_{c,z}}+\frac{\hbar^2k_\perp^2}{2m_{c,\perp}}+\beta_ck_zs.
\label{eq:EcCN}
\end{equation}
Here $k_z$ and $k_\perp$ are the components of the electron wavevector along the $z$ axis and in the plane perpendicular to it (reckoned from the $M$ or $M'$ point), $E_g$ is the bandgap energy and $\beta_c$ describes the spin-orbit splitting in the conduction band. The degeneracy of the valence band is completely lifted and in electron representation the dispersion of the topmost valence subband reads
\begin{equation}
  E_v=-\frac{\hbar^2k_z^2}{2m_{v,z}}-\frac{\hbar^2k_\perp^2}{2m_{v,\perp}}+(E_{k_z}-\Delta),
\label{eq:EvCN}
\end{equation}
where $E_{k_z}=\sqrt{\Delta^2+\beta^2k_z^2}$ with $\Delta$ and $\beta$ being the material parameters. The quantity $2\Delta>0$ determines the splitting between the valence subbands. The parameters $\beta_c$ and  $\beta$ change the signs at the transformation between the right and left handed modifications of Te. The wavefunction in the valence band is combined from the states $|J_z\rangle$ with spin projections $J_z=\pm3/2$ as
\begin{equation}
  \psi_v=C_{3/2}\left|3/2\right\rangle-C_{-3/2}\left|-3/2\right\rangle, \quad
 C_{\pm3/2}=\sqrt{\frac{E\mp\beta k_z}{2E}}.
\label{eq:psi_v}
\end{equation}
Equation~\eqref{eq:psi_v} shows that the valence band electron spin orientation is locked with the its wavevector, hence, the electric current in the valence band is directly related to the hole spin polarization. 
Equations~\eqref{eq:EcCN}---\eqref{eq:psi_v} have the same form in $M$ and $M'$ valleys of the energy spectrum.
In this particular material the hole spin and current fluctuations can not be separated.

We consider the experimentally relevant case of the $p$-doped Te~\cite{PhysRevB.93.045207} and assume that the light propagates along the
threefold rotation axis $z$.To demonstrate the effect, we consider the interband transitions between the topmost valence subband and the conduction band, as illustrated in Fig.~\ref{fig:Te}(b). At the microscopic level the contributions to dielectric tensor components from these transitions can be expressed as~\cite{agr_ginz}
\begin{multline}
  \delta\varepsilon_{xy}=- \delta\varepsilon_{yx}= \\
\frac{2\pi\i}{{\mathcal V}}{\sum_{\bm k}} \frac{\left|d^-(\bm k)\right|^2-\left|d^+(\bm k)\right|^2}{E_c-E_v-\hbar\omega}\left[f_h(\bm k)+f_h'(\bm k)\right],
\label{eq:epsilon}
\end{multline}
 where $f_h(\bm k)$ [$f_h'(\bm k)$] is the distribution function of holes in the topmost valence subband in the valley $M$ [$M'$], 
$d^\pm(\bm k)$ are the dipole moment matrix elements in $\sigma^\pm$ circular polarizations, and we have neglected the spin-orbit splitting of the conduction band because $\left|\beta\right|\gg\left|\beta_c\right|$. Hereafter $\mathcal V$ is used as the normalization volume. We note that the states in $M$ and $M'$ valleys are related by the time reversal symmetry only and the current induced contributions to $\varepsilon_{xy}$ from the two valleys have the same absolute value and sign. The quantity $p_{yxz}$, Eq.~\eqref{fluct}, describing the stochastic current induced Faraday rotation in Eqs.~\eqref{corr:F}, \eqref{eq:RMS} thus reads
\begin{equation}
\label{pyxz:Te}
p_{yxz}=\frac{2\pi |d_{cv}|^2\beta m_{v,z}}{\hbar |e|\Delta} \frac{1}{E_g -\hbar\omega}.
\end{equation}
Here $e$ is the electron charge, $d_{cv}$ is the interband momentum matrix element (which can be related with $m_{v,\perp}$ and $m_{c,\perp}$ within the two band model~\cite{averkiev1984:exp}). Hereafter we assume that the probe energy is below the absorption threshold, $\hbar\omega < E_g$, so that the probe beam propagates within the transparency frequency range. In derivation of Eq.~\eqref{pyxz:Te} we have considered only linear in $\beta$ contributions. Equation~\eqref{pyxz:Te} is valid for small detunings, $E_g - \hbar\omega \ll E_g$ and for not too high hole density, such that the hole Fermi energy $E_F \ll E_g - \hbar\omega, \Delta$. It is in agreement with Ref.~\cite{vorob1979optical} where the Faraday rotation caused by the homogeneous \emph{dc} current in Te was studied. 

It is instructive to introduce the parameter~\cite{Shalygin2012}
\begin{equation}
 \alpha=\frac{p_{yxz}\omega}{2cn_\perp}
\end{equation}
with $n_\perp$ being the refraction index in the plane perpendicular to the main crystal axis. It follows from Eq.~\eqref{fluct:F} that $\alpha$ is an analog of the Verdet constant for the current induced Faraday rotation, and can be measured experimentally. The mean square of the fluctuations of current induced Faraday rotation can be found using Eq.~\eqref{eq:RMS}.
For the estimation we take the values close to the parameters used in Ref.~\onlinecite{Shalygin2012}, namely, $\alpha=95~\mu$m/A, $\mathcal{L}=15$~mm, $N=5\cdot10^{16}$~cm$^{-3}$, $T=77$~K, $\mathcal{S}=\pi d^2/4$ with $d=1$~mm, and ${m_{v,z}}=0.08m_e$ with $m_e$ being the free electron mass.
For this set of parameters we obtain $\sqrt{\left\langle\delta\theta_F^2\right\rangle}\approx 5\times 10^{-2}~\mbox{mrad}$. 

\subsection{Chiral carbon nanotubes}\label{subsec:CNT}

\begin{figure}
  \centering\includegraphics[width=\linewidth]{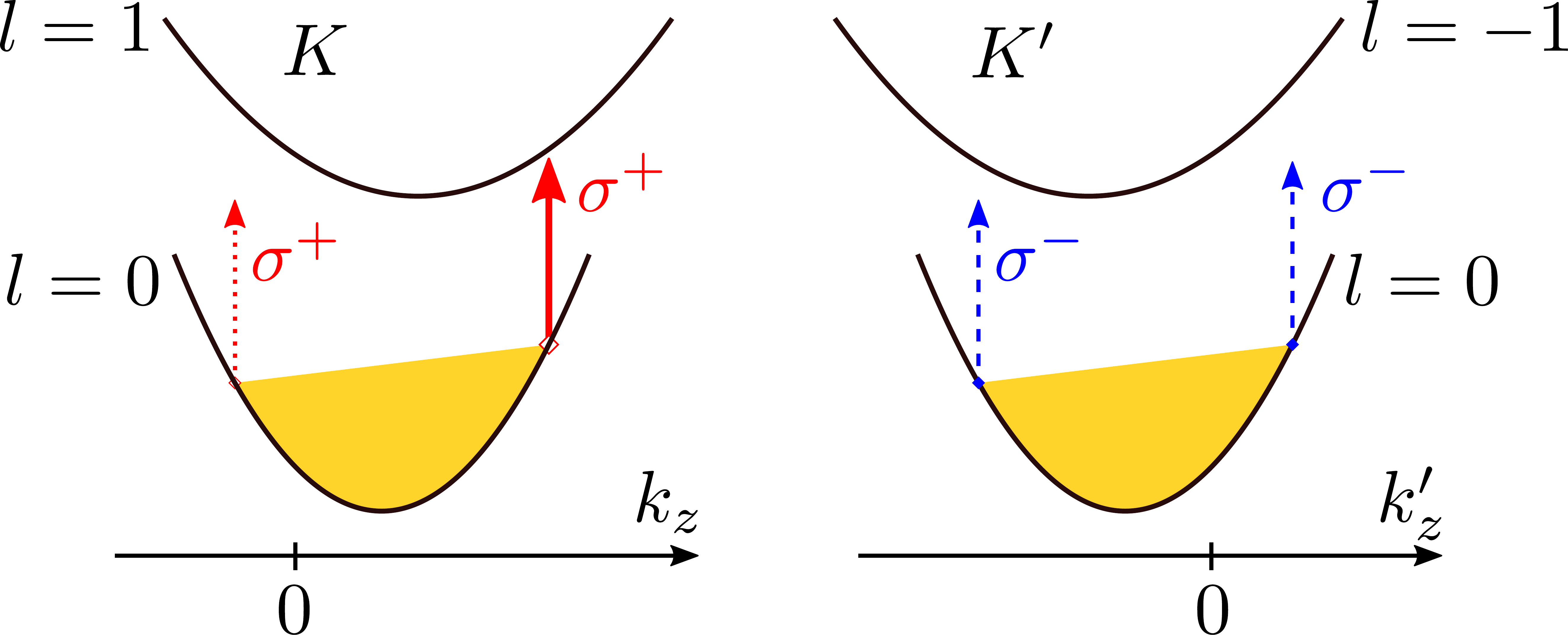}
  \caption{Electron dispersion in the ground and the first excited conduction subbands of the chiral carbon nanotube in the $K$ and $K'$ valleys, respectively. Solid and dashed arrows show the transitions with smaller and larger detunings, respectively, whose imbalance results in the Faraday rotation.}
  \label{fig:disp_CN}
\end{figure}

Another interesting gyrotropic system is an ensemble of chiral carbon nanotubes~\cite{iijima1991helical}. The single walled carbon nanotube can be conveniently characterized by the circumferential vector mapping the two dimensional graphene sheet to the cylinder circumference~\cite{ivchenko05a,saito1998physical}, $\bm L = n_a\bm a+ n_b \bm b$, where $n_a$, $n_b$ are the integers and $\bm a$, $\bm b$ are the basic vectors of the graphene lattice.
There are two series of conduction and valence subbands in carbon nanotubes stemming from the states in the two valleys $K$ and $K'$ of graphene. The dispersion relations for these subbands read
\begin{multline}
  E_{c,v}(l,k_z;K)=\pm\left(\frac{\Delta_l}{2}+\frac{\hbar^2 k_z^2}{2 m_l}+\beta_l k_z\right),\\
  E_{c,v}(l,k_z';K')=\pm\left(\frac{\Delta_l'}{2}+\frac{\hbar^2 k_z'^2}{2 m_l'}+\beta_l' k_z\right),
\label{eq:disp_CN}
\end{multline}
where the integer $l$ enumerates the subbands, $k_z$ ($k_z'$) is the wavenumber of the translational motion along the nanotube axis $z$ reckoned from the corresponding Brillouin zone edge, the remaining parameters such as the energy gaps $\Delta_l$ ($\Delta_l'$), effective masses $m_l$ ($m_l'$) and the constants $\beta_l$ ($\beta_l'$) can be expressed through the parameters of the graphene sheet, i.e. transfer integral and the lattice constant, and the angle between the vectors $\bm L$ and $\bm a$~\cite{ivchenko05a}.
Top and bottom signs in Eq.~\eqref{eq:disp_CN} correspond to the conduction and valence bands, respectively. 
The time reversal symmetry ensures that $E_{c,v}(l,k_z;K) = E_{c,v}(-l,-k_z,K')$ in Eq.~\eqref{eq:disp_CN}. The nanotube chirality results in the coupling between the translational motion along the nanotube axis and the electron angular momentum $l$ making the electron spectrum asymmetric in the $k$-space, see Fig.~\ref{fig:disp_CN}(a) and Ref.~\cite{ivch_spi} for details.

In order to simplify the derivation of the off diagonal components of the dielectric tensor responsible for the Faraday rotation we assume that the nanotubes form a three dimensional ensemble; the tubes in the ensemble are identical, have the same chirality 
$\nu = (n_a+n_b)~\mbox{mod}~3=1$,
and are oriented along $z$ axis. The nanotube chirality is most prominent for intersubband transitions for light propagating along the nanotube axis. These transitions are accompanied by the change of the subband state $l$ by $\pm 1$ depending on the incident light helicity (the interband transitions are considered in detail in Ref.~\cite{PhysRevB.72.195403}). Hence, we focus on the energy range of $l=0$ to $l=\pm 1$ intersubband transitions. We assume that the length of each nanotube is by far smaller than the wavelength of the incident light, therefore, we use the effective medium approximation
and present the dielectric tensor components analogously to Eq.~\eqref{eq:epsilon}
\begin{multline}
  \delta\varepsilon_{xy} = -\delta \varepsilon_{yx}= \\ 
  \frac{4\pi\i}{{\mathcal V}}   \sum_{k_z,j}\left[
\frac{\left|d(k_z)\right|^2 f_j'(k_z)}{E_c(-1,k_z;K')-E_c(0,k_z;K')-\hbar\omega} -
\right. 
\\
\left.
\frac{\left|d(k_z)\right|^2 f_j(k_z)}{E_c(1,k_z;K)-E_c(0,k_z;K)-\hbar\omega}\right].
\label{eq:epsilon_CN}
\end{multline}
Here the extra factor $2$ as compared with Eq.~\eqref{eq:epsilon} results from the summation over the spin states, $j$ enumerates the nanotubes in the ensemble, $f_{j}(k_z)$ [$f_j'(k_z)$] is the electron distribution function in the ground conduction subband of the valley $K$ [$K'$], and the occupation of excited subbands is neglected. The squared absolute value of the dipole moment operator for the intersubband transition reads $|d(k_z)|^2=d_0^2k_z^2$, where $d_0$ is a parameter~\cite{ivch_spi}.

Equation~\eqref{eq:epsilon_CN} shows that there are two contributions to the off diagonal components of the dielectric tensor. First, the imbalance of the occupancies in $K$ and $K'$ valleys results in the preference of the transitions in $\sigma^+$ or $\sigma^-$ polarizations yielding the valley polarization induced Faraday rotation:
\begin{equation}
\label{CNT:valley:0}
\delta \varepsilon_{xy} = - \delta \varepsilon_{yx} = - \mathrm i p_v n_v.
\end{equation}
Here $n_v = 2{\mathcal V^{-1}} \sum_{k_z,j} [f_j(k_z)-f_j'(k_z)]$ is the valley population imbalance in the nanotube ensemble and
\begin{equation}
\label{pv}
  p_{v}=\frac{2\pi\left|d(k_F)\right|^2}{\delta},
\end{equation}
where $\delta=\Delta_0/2 - E_F/2 - \hbar\omega$ and $E_F=\hbar^2k_F^2/2m_0$ with $k_F$ being the Fermi wavevector and $E_F$ being the Fermi energy.
Equation~\eqref{pv} is derived for the degenerate electrons, $k_BT\ll E_F,\delta$, with not too high density where $E_F,\delta \ll \Delta_0$, and we have taken into account that $\Delta_1=2\Delta_0$ and $m_1=2m_0$. Corresponding Faraday rotation noise power spectrum reads
\begin{equation}
\label{CNT:valley}
(\delta \theta_F^2)^{v}_\Omega = \left(\frac{\omega{\mathcal{L}}}{2cn}\right)^2  p_v^2 \left({\delta n_v^2}\right)_{\Omega},
\end{equation}
where $\left({\delta n_v^2}\right)_{\Omega}$ is the power spectrum of the valley population imbalance in the nanotube ensemble~\cite{Note3}. Second, the asymmetry of the conduction subbands energy spectrum gives rise to the current induced Faraday rotation with 
 \begin{equation}
\label{pyxz:CNT}
  p_{yxz}=-\frac{2\pi\beta_0m_0}{e\hbar}\frac{|d(k_F)|^2}{\delta^2},
\end{equation}
where we have used the relation $\beta_1=2\beta_0$.

In order to estimate the same time mean square of the Faraday rotation angle we make use of the relations
\begin{subequations}
\begin{align}
&\int (\delta j_z^2)_\Omega d\Omega = \frac{4e^2}{{\mathcal V}^2}\sum_{k_z} \left(\frac{\hbar k_z}{m_0}\right)^2 f_{k_z}(1-f_{k_z}) =\frac{2\pi k_B T}{\mathcal V} \frac{Ne^2}{m_0},\\
&\int (\delta n_v^2)_\Omega d\Omega = \frac{4e^2}{{\mathcal V}^2}\sum_{k_z} f_{k_z}(1-f_{k_z}) = \frac{{\pi}k_B T}{\mathcal V} {\frac{N}{E_F}}.
\end{align} 
\end{subequations}
where $f_{k_z}$ is the equilibrium distribution function, $N$ is the concentration of electrons in the ensemble.
These equations express the single time fluctuations of the current and of the valley population imbalance.

Interestingly,  the valley polarization fluctuations induced Faraday rotation noise, Eqs.~\eqref{pv}, \eqref{CNT:valley}, and the current fluctuations induced Faraday rotation noise, Eqs.~\eqref{pyxz:CNT}, \eqref{corr:F}, demonstrate quite different dependence on the detuning and, hence, on the probe frequency $\omega$.  This is because $p_v$ is inversely proportional to the detuning, $\delta$, while $p_{yxz}$ is inversely proportional to $\delta^2$.  The different dependence of $\langle \theta_F^2\rangle$ for two mechanisms on the detuning allows one, in principle, to separate these two contributions in the experiments similar to those described in Ref.~\cite{PhysRevLett.110.176601}.

Another way to separate the contributions due to current and valley polarization noise is their different power spectra as functions of the noise frequency $\Omega$. The different dependence on $\Omega$ is due to the fact that the timescale of valley fluctuations, that is the intervalley scattering time $\tau_v$, is, as a rule, strongly different from that of the current fluctuations controlled by shorter momentum relaxation time $\tau$ or the time of flight for ballistic nanotubes. Moreover, weak electrical connections between the nanotubes in the ensemble may result in the specific frequency dependence of $\sigma(\Omega)$, particularly with a power law feature at $\Omega\to 0$~\cite{Zvyagin}.

For the estimations we take $E_F=10$~meV, $L=63$~nm,
$\beta_0=-2.3$~meV$\cdot$nm,
 $T=10$~K, $\mathcal{S}=50~\mu\rm m^2$,
two dimensional concentration of nanotubes $\mathcal{L}N=2.5\cdot10^{11}$~cm$^{-2}$
and make use of the standard expressions introduced in Ref.~\cite{ivch_spi} to derive $m_0,\Delta_0$ and $d_0$. At the detuning $\delta=2$~meV the mean square fluctuation caused by valley polarization fluctuations amounts to $\sim10^{-3}$~mrad and by current fluctuations to $\sim5\cdot10^{-5}$~mrad. These contributions are of the same order at smaller detuning $\delta=0.05$~meV, but in this case the fluctuations of higher powers of the current density or the correlations between the valley polarization and the current, if any, may play a role.

\begin{figure*}
  \centering
  \includegraphics[width=0.8\textwidth]{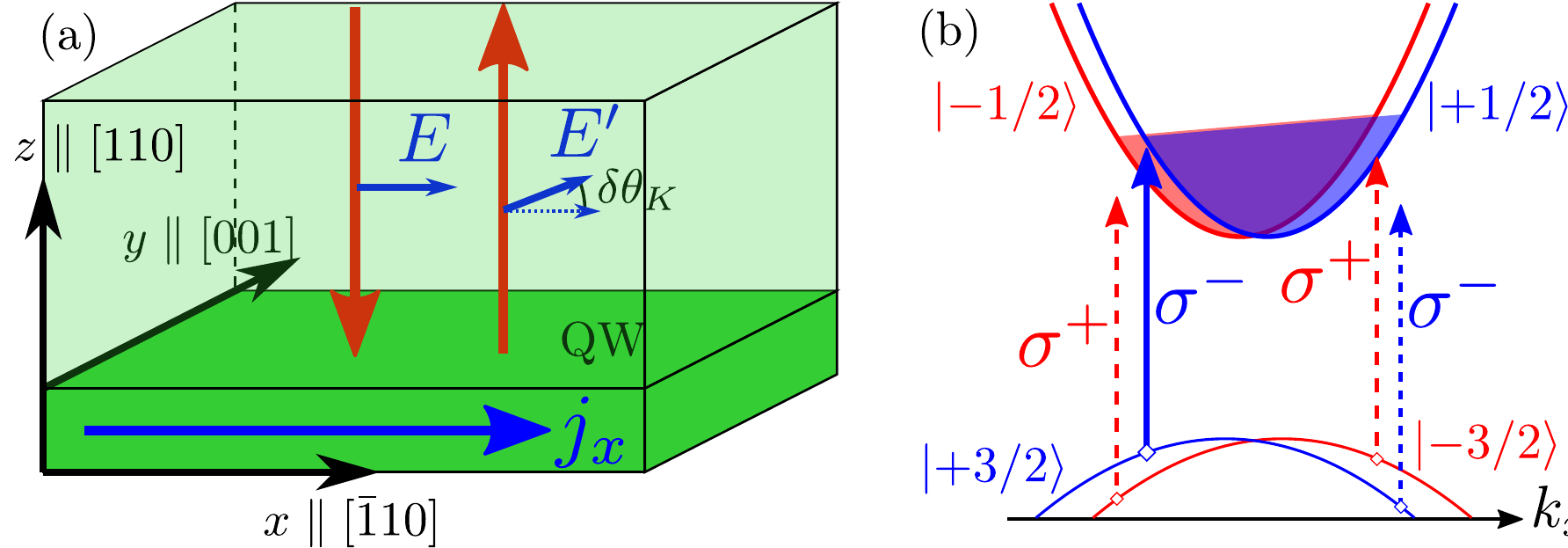}
  \caption{(a) Sketch of the polarization conversion effect in the reflection geometry for $[110]$ grown quantum well. (b)~Illustration of the band structure, 
 solid and dashed arrows show the transitions with smaller and larger detunings, respectively, whose imbalance results in the Faraday rotation.
Filled regions denote occupied states in the conduction band.
}
  \label{fig:QW}
\end{figure*}

Above we considered an ensemble of identical nanotubes. We note that in ensemble of nanotubes with different chiralities and equal numbers of $\nu=+1$ and $\nu=-1$ nanotubes, the gyrotropy is absent on average, but the current and valley polarization fluctuations induced stochastic Faraday rotation is still present.  


\subsection{Zincblende lattice quantum wells grown along the $[110]$ axis}\label{subsec:110}

Although bulk zincblende lattice semiconductors like GaAs are not gyrotropic, the QW structures grown from these materials demonstrate gyrotropy~\cite{2016arXiv160701226K}. In particular, quantum wells grown along $z\parallel[110]$ direction  with symmetric heteropotential have the $C_{2v}$ point symmetry group. In the coordinate frame with the in plane axes $x\parallel[\bar110]$, $y\parallel[001]$ the twofold rotation axis $C_2 \parallel y$ lies in the intersection of reflection planes $(xy)$ and $(yz)$, Fig.~\ref{fig:QW}(a). Hence, the in plane $x$-component of a vector, i.e., current density $j_x$, transform like $z$-component of a pseudovector. As a result, in the plane current fluctuations give rise to the noise of the Faraday or Kerr rotation for the probe beam incident along the structure normal~$z$. The origin of this effect is the spin orbit interaction, which allows for the terms in the effective Hamiltonian (see Refs.~\cite{dyakonov86,PhysRevB.85.205307} and references therein)
\begin{equation}
\label{HSO:110}
\mathcal H_{SO} = \beta_e\sigma_z^{(e)} k_x^e+\beta_h \sigma_z^{(h)} k_x^h,
\end{equation}
where $\beta_e$ ($\beta_h$) is the spin splitting parameters for the electron (hole), $\sigma_z^{(e)}$ ($\sigma_z^{(h)}$) is the spin (pseudospin) $z$-component Pauli matrix acting in the basis of $\pm 1/2$ electron ($\pm 3/2$ heavy-hole) spin states, $k_x^e$ ($k_x^h$) is the $x$-component of the electron (hole) wavevector.
In what follows, we consider the reflection geometry, as sketched in Fig.~\ref{fig:QW}(a), and calculate the Kerr rotation noise caused by the electric current fluctuations.

Optical response of two dimensional structures can be conveniently characterized by the matrix of reflection coefficients, $r_{\alpha\beta}$, which links the Cartesian components of the electric field in the incident and reflected waves. By analogy with Eqs.~\eqref{eq:epsilon} and~\eqref{eq:epsilon_CN} for the dielectric tensor, the off diagonal components of the reflection coefficients matrix read~\cite{averkiev07,glazov:review}
\begin{equation}
  r_{xy}=-r_{yx}=-\frac{2\pi\left|d\right|^2\omega}{c{\mathcal S}}\sum_{\bm k,s}\frac{s[1-f_{s}(\bm k)]}{E_g+\frac{\hbar^2k^2}{2\mu}-2s\beta k_x-\hbar\omega}.
\label{eq:rpm}
\end{equation}
Here we consider $n$-doped quantum well and focus on the spectral range of interband transitions [see Fig.~\ref{fig:QW}(b)], $f_{s}(\bm k)$ are the distribution functions of the electrons with $z$ spin component $s=\pm1/2$, $E_g$ is the band gap including the size quantization energies, $\beta=\beta_e- \beta_h$, $\mu$ is the reduced mass of electron and hole, $\mathcal S$ is taken to be the normalization area, and  $d$ is the dipole moment matrix element between the size quantized conduction and valence band electron states. The latter is given by~\cite{ivchenko05a}
\begin{equation}
  d=d_{cv}\int C_c^*(z)C_v(z) d z,
\end{equation}
where $d_{cv}$ is the dipole moment matrix element between conduction and valence band Bloch amplitudes at the $\Gamma$ point, $C_c(z)$ and $C_v(z)$ are the envelope functions at $\bm k=0$ in the conduction and valence bands, respectively. The Kerr rotation angle of the reflected light polarization plane can be expressed as~\cite{zhu07,PhysRevB.92.115305}
\begin{equation}
  \theta_K=\frac{r_{yx}}{r_0},
\end{equation}
where $r_0$ is the background reflection coefficient, which is usually determined by the cap layer, for simplicity it is assumed to be real. Here, as before, we neglect the regular optical anisotropy of the structure.
The off diagonal transmission coefficients have the same form as $r_{xy}$, $r_{yx}$ in Eq.~\eqref{eq:rpm}.

Similarly to the chiral carbon nanotubes considered in Sec.~\ref{subsec:CNT} where the stochastic Faraday rotation is contributed both by current and valley noise, in semiconductor quantum wells, in addition to the current induced contribution, which is in the main focus of our paper, the spin polarization also contributes to the Faraday and Kerr effects. The spin polarization effect has been already studied in detail theoretically and experimentally, see, e.g., Refs.~\cite{glazov:review,PhysRevB.89.081304} and references therein. Correspondingly,
\begin{equation}
  r_{xy}=P_ss_z + P_{xyx} j_x,
\end{equation}
where $s_z = \sum_{\bm k} \left[f_{1/2}(\bm k) - f_{-1/2}(\bm k)\right]/(2\mathcal S)$ is the spin density and
\begin{equation}
  P_s=\frac{2\pi\omega\left|d\right|^2}{c(E_g - \hbar\omega)},
\end{equation}
\begin{equation}
\label{Pxyx}
  P_{xyx}=\frac{\pi\omega\left|d\right|^2\beta m_c}{ec\hbar(E_g - \hbar\omega)^2},
\end{equation}
with $m_c$ being the electron effective mass, and, as above, we neglected the shift of the absorption edge caused by the state filling effect as well as the many body effects~\cite{averkiev07}. In two dimensional system instead of Eq.~\eqref{eq:Faraday_sigma} we have
  \begin{equation}
    (\delta \theta_F^2)_\Omega = \frac{2k_B T }{\mathcal S} \left(\frac{P_{xyx}}{r_0}\right)^2\Re\, \sigma_{xx}^{(2d)}(\Omega),
\label{eq:Faraday_sigma:2d}
  \end{equation}
where $\sigma_{xx}^{(2d)}(\Omega)$ is the two dimensional \emph{ac} conductivity given by the Drude formula Eq.~\eqref{eq:Drude} with $N$ being the two dimensional electron density.  The root mean square of the current noise induced Kerr rotation noise amounts to $\sqrt{\left\langle\delta\theta_K^2\right\rangle}\sim0.1~\mbox{mrad}$ for the parameters
$E_g=1.5$~eV, interband matrix element $E_P=28.8$~eV,
 $m_c=0.067 m_e$, $\beta=10$~meV$\cdot$nm~\cite{PhysRevLett.104.066405,PhysRevB.89.075430},  $r_0=0.5$, $E_g-\hbar\omega=2$~meV, $N=3\cdot10^{11}$~cm$^{-2}$, $\mathcal{S}=7~\mu$m$^2$, $T=77$~K. At the same conditions, the root mean square of the spin noise induced Kerr angle is just somewhat larger, $\sim 0.2~\mbox{mrad}$~\cite{Note4}.

Note, that in multiple quantum well structures with sufficiently thick barriers where both the electron tunnelling between the wells and the Coulomb interwell interaction are suppressed, the current induced noise power increases linearly with the number of quantum wells in the structure.

\subsection{Zincblende lattice quantum wells grown along the $[001]$ axis}\label{subsec:001}

For the normal light incidence the current induced Faraday and Kerr effects are symmetry forbidden in zincblende quantum wells grown along the cubic axis, $z\parallel [001]$. However, these quantum wells are gyrotropic as well, see Ref.~\cite{2016arXiv160701226K} and references therein. Here the current induced rotation of the polarization plane is possible at the oblique incidence only. To illustrate this phenomena we consider the case of a $[001]$ grown quantum well with the dominant structural inversion asymmetry.

\begin{figure}
  \centering
  \includegraphics[width=\linewidth]{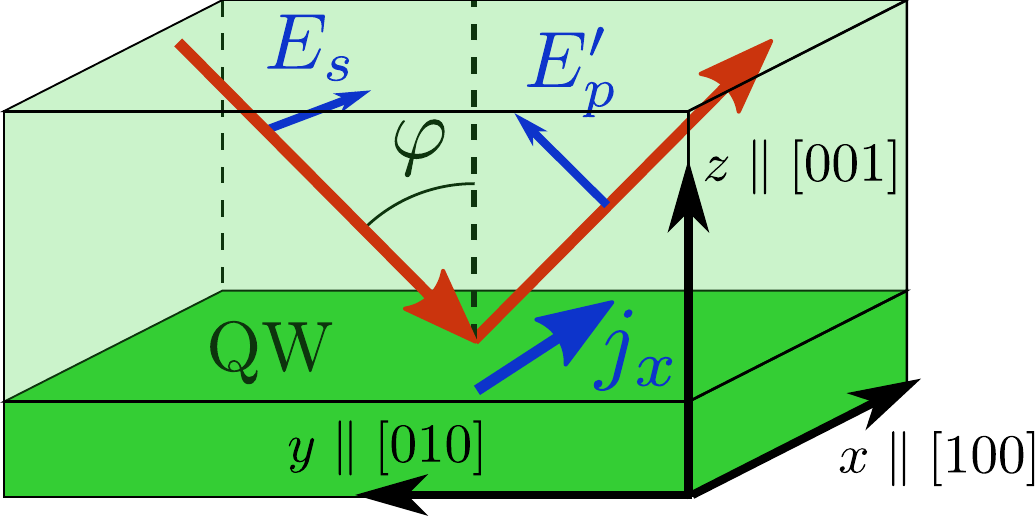}
  \caption{Illustration of polarization conversion effect for $[001]$ grown quantum well under oblique light incidence.}
  \label{fig:QW001}
\end{figure}

Let us consider the $s$-polarized light incident upon the quantum well, let $(yz)$  be the incidence plane and $\varphi$ be the angle of incidence inside the structure, Fig.~\ref{fig:QW001}(b), the in plane axes are $x\parallel [100]$, $y\parallel [010]$. The Kerr rotation is described by the off diagonal reflection coefficients $r_{ps}=-r_{sp}\propto j_x$~\cite{Note5}. Taking into account the heavy-light hole mixing via the off diagonal element $H \propto k_z(k_x-\mathrm i k_y)$ of the Luttinger Hamiltonian~\cite{ivchenko05a} in the first order we present the electron wavefunction in the valence band with the in plane wave vector $\bm k$ in the form~\cite{merkulov90_eng}
\begin{multline}
  \Psi_{\pm3/2}(\bm k)=\\\frac{e^{\i \bm{k \rho}}}{\sqrt{\mathcal S}}\left[C_{v}(z)|\pm3/2\rangle \mp \mathrm i (k_x\pm\i k_y)a S_{v}(z)|\pm1/2\rangle\right],
\end{multline}
where $a$ is the quantum well width, 
$\bm \rho$ is the in plane position vector, $|\pm 3/2\rangle, |\pm 1/2\rangle$ are the Bloch amplitudes of the $\Gamma_8$ valence band, and $C_v(z),~S_v(z)$ are the envelope functions describing size quantization. Note that in quantum wells with asymmetric heteropotential these functions does not possess a certain parity with respect to $z\to -z$ reflection. Making use of the selection rules for the interband optical transitions we obtain
\begin{equation}
r_{ps}=\frac{2\pi\omega}{c\mathcal S}\sum_{\bm k,s}\frac{f_sd_\perp\left[d_z^*\left(k_x-2\i s k_y\right)a\sin\varphi-2d_\perp^* s\cos\varphi\right]}{E_g+\frac{\hbar^2k^2}{2\mu}-\hbar\omega},
\label{eq:rps}
\end{equation}
where
\begin{subequations}
\begin{align}
  {d_\perp}=-\frac{d_{cv}}{\sqrt{2}}\int {C_c^*(z) C_v(z)} d z,\\
  {d_z}=\sqrt{\frac{2}{3}}d_{cv}\int {C_c^*(z) S_v(z)} d z,
\end{align}
\end{subequations}
and the other notations are the same as in previous subsection. The fact that simultaneously both $d_z$ and $d_\perp$ are nonzero is related with the structural inversion asymmetry of the considered quantum well.

One can see from Eq.~\eqref{eq:rps}, that apart from the well studied spin induced Faraday rotation, described by the last term in the square brackets in Eq.~\eqref{eq:rps}, at oblique incidence there are two additional contributions to the polarization conversion effect
  \begin{equation}
    r_{ps}=P_{psx}\left(j_x-2\i J_y^z\right),
  \end{equation}
with $J_y^z$ being the spin current density along $y$ axis. The fact that the spin current contributes to polarization conversion is expected, because $j_x$ induces $J_y^z$ as a result of the spin-Hall effect~\cite{dyakonov_book,Note6}.
The calculation yields
\begin{equation}
\label{Ppsx}
P_{psx} = \frac{2\pi \omega m_c a}{ec\hbar}\frac{d_\perp d_z^*\sin\varphi}{E_g - \hbar\omega},
\end{equation}
which is derived under the same assumptions as Eq.~\eqref{Pxyx}. The Kerr rotation angle reads 
$\theta_K=\Re (r_{ps}/r_{ss})$, where $r_{ss}$ is the reflection coefficient for $s$ polarized light.
Note that accounting for the phase associated with the light propogation through the cap layer the $r_{ss}$ becomes complex, which allows for optical detection of spin current fluctuations.
The power spectrum of the Kerr rotation noise related to electric current fluctuations is given by Eq.~\eqref{eq:Faraday_sigma:2d} with the replacement of $P_{xyx}$ by $P_{psx}$. The estimates show that $\sqrt{\left\langle\delta\theta_K^2\right\rangle}\sim0.1~\mbox{mrad}$ at $a=100$~\AA, the incidence angle $\varphi=\pi/3$, $r_{ss}=0.5$ and other parameters as in Sec.~\ref{subsec:110}.

\section{Discussion}\label{sec:discussion}

In gyrotropic systems the $\sigma^+$ and $\sigma^-$ polarized light differently interacts with the electrons with the given momentum. The fluctuations of the current in the sample lead to the redistribution of the electrons in the momentum space, and therefore result in fluctuations of the light refraction indices for the two circular polarizations, $n_\pm$. Hence, the stochastic Faraday or Kerr rotation is induced in gyrotropic systems due to the electric current noise.

Just like in any system, the electron spin fluctuations ${\delta \bm s}$ manifest themselves as an additive to Eq.~\eqref{fluct:F} contribution to the Faraday rotation angle, $\delta \theta_F'=(\omega/2cn) g_{\alpha\beta\gamma} \int d\bm r \delta {s}_\gamma(t,\bm r)/\mathcal S$. Here the third rank pseudotensor $g_{\alpha\beta\gamma} = - g_{\beta\alpha\gamma}$ describes spin induced optical activity, $\varepsilon_{\alpha\beta} = \mathrm i g_{\alpha\beta\gamma} {s}_\gamma$~\cite{ll8_eng,aronovivchenko:eng,PhysRevB.85.195313}. As a result, the current noise in the Faraday rotation fluctuation spectrum, Eq.~\eqref{corr:F}, is superimposed on the spin noise spectrum. As a rule, the typical timescales of spin dynamics (spin relaxation time $\tau_s$, spin precession period in the external magnetic field $2\pi/\Omega_L$) are much longer as compared with the timescales of the orbital dynamics of the electrons given by the momentum relaxation time $\tau_p$ or the cyclotron period $2\pi/\Omega_C$ due to relative weakness of the spin orbit coupling in many semiconductor systems. Hence, the spin and current fluctuations can be readily separated in the experiment~\cite{Note7}. Moreover, the gyrotropy of the considered systems allows for linear relations between the spin and current (spin galvanic and spin orientation by the electric field effects), this makes spin and current fluctuations interrelated. The analysis of such cross correlation noise and its manifestations is beyond the scope of the present paper. For the same reasons we do not address the cross correlations between, e.g., the electric current and valley polarization fluctuations in carbon nanotubes.

The manifestation of the current noise in the Faraday or Kerr rotation fluctuations allows, in principle, for the contactless measurement of the conductivity of the system. It may serve as an alternative to the studies of the transport response function in the light scattering experiments~\cite{platzman1973solid,springerlink:10.1007/BF02724353,Bareikis}. The link between the two approaches is similar to the relation between the spin noise spectroscopy and the spin-flip Raman scattering spectroscopy, see Refs.~\cite{gorb_perel,Glazov:15} for details.


\section{Conclusion}\label{sec:conclusion}

To conclude, we have demonstrated that in gyrotropic systems the electric current fluctuations result in the stochastic rotation of the polarization plane of the linearly polarized light propagating through the system. Just like spin fluctuations, the current noise results in the noise of the Faraday or Kerr rotation effects. Hence, in gyrotropic materials one can access the electric current correlation function, and, particularly, the \emph{ac} conductivity by means of optical noise spectroscopy. The microscopic theory of the effect has been developed for bulk tellurium, ensembles of carbon nanotubes, and two-dimensional systems, such as GaAs quantum wells grown along different crystallographic directions. The estimates show that among the studied systems the effect is strongest in quantum well structures.

\acknowledgements

We thank E.\,L. Ivchenko for fruitful discussions and S.\,G. Smirnov for technical support. Partial  support from RFBR (projects 15-02-06344 and 15-32-20828), the Dynasty Foundation and RF President Grant No. SP-643.2015.5 is gratefully acknowledged.



%

\end{document}